\documentclass[twocolumn,a4paper,10pt]{revtex4}
\usepackage{graphicx,psfrag}
\usepackage{amsmath,amssymb}
\usepackage{multirow}
\pdfoutput=1
\begin{document}
\title{Five models for myosin V}
\author{Andrej Vilfan}
\affiliation{J. Stefan Institute, Jamova 39, 1000
  Ljubljana, Slovenia}
\begin{abstract}
Myosin V was the first discovered processive motor from the myosin family. It
has therefore been subject of a number of mechanical, kinetic, optical and
structural studies and now belongs to the best characterised motor proteins.
This effort has been accompanied by a number of different theoretical
models. In this article we give an overview of them and discuss what they have
in common and where the open questions are.  The latter include the existence
of sub-steps, the process that limits the run length, the nature of backward
steps, the flexibility of the lever arm and the state of the lead head.
\end{abstract}
\maketitle

\section{Introduction}
Myosin V is a two-headed processive motor protein from the myosin superfamily
\cite{Kieke.Titus2003}, involved in different forms of intracellular transport
\cite{Reck-Peterson.Mercer2000,Langford2002}. In 1999 it was shown to be the
first processive motor from the myosin superfamily \cite{Mehta.Cheney1999}.
Processivity means that a single molecule can move over a long distance
(typically a micron) along an actin filament without dissociating.  This
discovery came as a surprise, because processivity was only known for
microtubule-based motors (kinesin, ncd and cytoplasmic dynein) before.  The
second feature that makes myosin V unconventional is its long step size around
$35\,{\rm nm}$ \cite{Mehta.Cheney1999}. After myosin V, other processive
myosins have been discovered (VI \cite{Rock.Sweeney2001}, VIIa
\cite{Yang.Sellers2006}, XI \cite{Tominaga.Oiwa2003}, X \cite{Rock2008} and
according to some evidence IXb \cite{Post.Mooseker2002}), but the amount of
knowledge gained on myosin V has made it one of the best, if not the best
understood motor protein.

The experiments have characterised it mechanically (usually using optical
tweezers)
\cite{Mehta.Cheney1999,Rock.Spudich2000,Rief.Spudich2000,Veigel.Molloy2002,Purcell.Sweeney2002,Clemen.Rief2005,Gebhardt.Rief2006,Uemura.Ishiwata2004},
biochemically
\cite{De_La_Cruz.Sweeney1999,De_La_Cruz.Ostap2000a,De_La_Cruz.Ostap2000b,Yengo.Sweeney2002},
optically \cite{Ali.Ishiwata2002,Forkey.Goldman2003,Yildiz.Selvin2003} and
structurally
\cite{Walker.Knight2000,Burgess.Trinick2002,Wang.Sellers2003,Coureux.Houdusse2003}.
These studies have shown that myosin V walks along actin filaments in a
hand-over-hand fashion \cite{Yildiz.Selvin2003,Warshaw.Trybus2005} with an
average step size of about 35 nm, roughly corresponding to the periodicity of
actin filaments
\cite{Mehta.Cheney1999,Rief.Spudich2000,Walker.Knight2000,Veigel.Molloy2002,Ali.Ishiwata2002},
a stall force of around 2.5 pN \cite{Rief.Spudich2000} and a run length of a
few microns \cite{Rief.Spudich2000,Sakamoto.Sellers2003,Baker.Warshaw2004}.
Under physiological conditions, ADP release has been identified as the time
limiting step in the duty cycle \cite{De_La_Cruz.Sweeney1999,Rief.Spudich2000},
which means that the motor spends most of its time waiting for ADP release from
the trail head.

Myosin V contains a head domain, which has a high degree of similarity with
other myosins like myosin II (muscle myosin).  However, there are important
differences in the kinetics of the ATP hydrolysis cycle -- notably the slow
release of ADP, which is necessary for maintaining processivity.  The head is
connected with a lever arm (also called neck), consisting of an alpha helix,
surrounded by six light chains.  This is followed by the tail domain, which is
responsible for the dimerisation of the molecule and for cargo binding.

Meanwhile the literature on myosin V counts about a dozen of reviews,
concentrating on different aspects of myosin V, such as its cellular function
and phylogeny
\cite{Titus1997,Provance.Mercer1999,Reck-Peterson.Mercer2000,Langford2002,Kieke.Titus2003,Desnos.Darchen2007,Trybus2008,Sellers.Weisman2008},
regulation \cite{Taylor2007,Trybus2008}, structure \cite{Sweeney.Houdusse2004}
and stepping mechanism
\cite{Mehta2001,Tyska.Mooseker2003,Vale2003b,Olivares.De_La_Cruz2005,Sellers.Veigel2006,Schmitz.Veigel2006,Sellers.Weisman2008}.
It is not the purpose of this article to review all important findings on
myosin V in general, but rather to concentrate on theoretical models and those
experimental results they aim at explaining.  More specifically, we will
critically review 5 existing theoretical models
\cite{Kolomeisky.Fisher2003,Vilfan2005,Lan.Sun2005,Skau.Turner2006,Wu.Karplus2007}
and discuss which of them is in agreement with which of the experimental
results and where the discrepancies are.

The structure of this paper is as follows.  In Section \ref{sec:mainfindings}
we summarise those experimental results that are most relevant for theoretical
models.  Section \ref{sec:open-questions} describes some of the most important
open questions in the field -- particularly those that need cooperation between
theory and experiment to be resolved.  Section \ref{sec:theoretical-models}
then reviews five theoretical models for the stepping kinetics and mechanics of
myosin V.  Finally, we give a brief summary and try to project how the
different approaches might converge in near future.

\section{Main findings}
\label{sec:mainfindings}

\begin{figure}
\begin{flushleft}
  \includegraphics{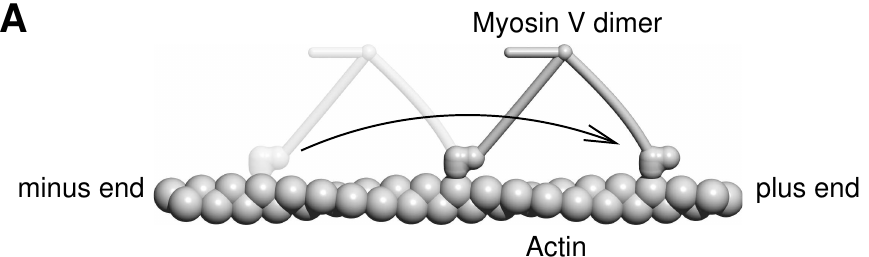}\\
  \includegraphics{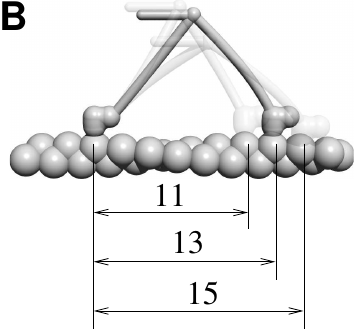}
\end{flushleft}
\caption{(A) Myosin V steps in a hand-over-hand fashion along actin filaments.
  (B) The most frequent step size is 13 actin subunits, but it can also be 11
  and 15 subunits occur}
\label{fig:intro}
\end{figure}

\begin{figure}[t]
  \begin{center}
    \includegraphics{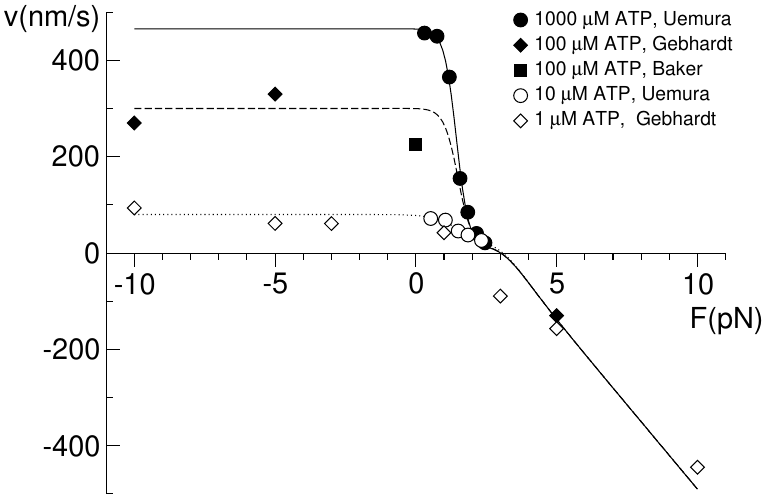}
  \end{center}
  \caption{Force-velocity relation of myosin V for different ATP
    concentrations.  Data points are taken from
    Refs.~\cite{Uemura.Ishiwata2004,Baker.Warshaw2004,Gebhardt.Rief2006}. The
    lines are guide to the eye. The relationship shows a rather flat region
    between $-10$pN and 1pN.  In that region, the velocity as a function of 
    ATP concentration follows a Michaelis-Menten-like relation.  This region is
    followed by an abrupt drop between 1pN and 2pN. Above the stall force, the
    motor is pulled backwards with a velocity that increases roughly linearly
    with load and is independent of the ATP and ADP concentration.}
  \label{fig:forcevelocity}
\end{figure}
The main findings gained with optical tweezers, fluorescent microscopy and
chemical kinetics can be summarised as follows:
\begin{enumerate}
\item Myosin V walks in a \emph{hand-over-hand fashion} along actin filaments
  \cite{Yildiz.Selvin2003,Warshaw.Trybus2005}, as shown in
  Fig.~\ref{fig:intro}A{}.  The lead head performs free Brownian motion around
  the common hinge while searching for the new binding site
  \cite{Dunn.Spudich2007,Shiroguchi.Kinosita2007}.
\item The \emph{step size} roughly corresponds to the actin periodicity of 35nm
  \cite{mehta99}.  The most precise way to determine the \textit{average} step
  size is by measuring the pitch of the slightly helical motion of the stepping
  myosin V molecule around the actin filament \cite{Ali.Ishiwata2002}.  The
  \textit{distribution} of step sizes can be determined from EM images
  \cite{Walker.Knight2000}.  The study shows the highest probability for 13
  subunit steps (36nm), followed by 15 subunit and 11 subunit steps (Fig.~\ref{fig:intro}B).
\item The \emph{force velocity relation}, shown in
  Fig.~\ref{fig:forcevelocity}, reveals the following phases:
  \begin{itemize}
  \item Under saturating ATP concentrations (in the millimolar range), the
    velocity is about 400nm/s and is nearly constant up to loads of about 1pN
    \cite{mehta99,Uemura.Ishiwata2004}.

    For higher loads, the velocity drops rather sharply and flattens near
    stall \cite{mehta99,Uemura.Ishiwata2004}.

    Above the stall force, when the motor is pulled backwards by the load, its
    velocity increases roughly linearly with the applied load; it reaches
    around $-150$nm/s \cite{Gebhardt.Rief2006} to $-250$nm/s \cite{Clemen.Rief2005}
    under 5pN load and around $-700$nm/s under 10pN \cite{Gebhardt.Rief2006}.

    Forward loads only lead to a slight increase in velocity
    \cite{Gebhardt.Rief2006,Clemen.Rief2005}.

  \item Under low ATP concentrations, the velocity becomes proportional to the
    ATP concentration and is independent of load between $-10$pN and 2pN
    \cite{Gebhardt.Rief2006,mehta99,Uemura.Ishiwata2004}.  For super-stall
    loads, the velocity becomes independent of the ATP concentration.

  \item The stall force of myosin V has been reported as 3pN
    \cite{Mehta.Cheney1999}, 2.5-3pN \cite{Uemura.Ishiwata2004}, up to 2.5pN
    \cite{Cappello.Prost2007}, but also lower values around 1.7pN
    \cite{Clemen.Rief2005}.  Because of short run lengths, some care has to be
    taken when determining the force-velocity relation close to stall force, as
    pointed out by Kolomeisky \emph{et al.}\ for kinesin
    \cite{Kolomeisky.Popov2005}.

  \item Added ADP has a similar effect on the force-velocity curve as a reduced
    ATP concentration
    \cite{Rief.Spudich2000,Baker.Warshaw2004,Uemura.Ishiwata2004}.  Added
    inorganic phosphate in high concentrations (40mM) causes a less
    significant, but detectable velocity decrease
    \cite{Baker.Warshaw2004,Dunn.Spudich2007}.
  \end{itemize}
\item The second important source of data is the degree of processivity,
  usually measured with the average \emph{run length}, i.e., the average
  distance a single myosin V motor moves along an actin filament before
  dissociating from it.  There is still some discrepancy with regard to the run
  length of unloaded motors. The first measurement was carried out by Mehta
  \emph{et al.}\ \cite{Mehta.Cheney1999}, who estimated it as $2\,\mu \rm m$.
  Veigel \emph{et al.}\ \cite{Veigel.Molloy2002} obtained $2.4\,\mu \rm m$, Ali
  \emph{et al.}  \cite{Ali.Ishiwata2002} $2.1\,\mu\rm m$, Baker \emph{et al.}
  \cite{Baker.Warshaw2004} measure values around $1\,\mu \rm m$, while the data
  of Clemen \emph{et al.}\ \cite{Clemen.Rief2005} are closer to 300nm.  A
  possible explanation for this apparent discrepancy could lie in the fact that
  actin filaments were attached to the glass surface in the latter experiment,
  which possibly hindered the motor in its natural helical motion around the
  filament and caused a premature dissociation.  Therefore, one can accept run
  length values around $2\,\rm \mu m$ for unconstrained motion along actin
  filaments.

  \begin{figure*}
    \begin{center}
      \includegraphics{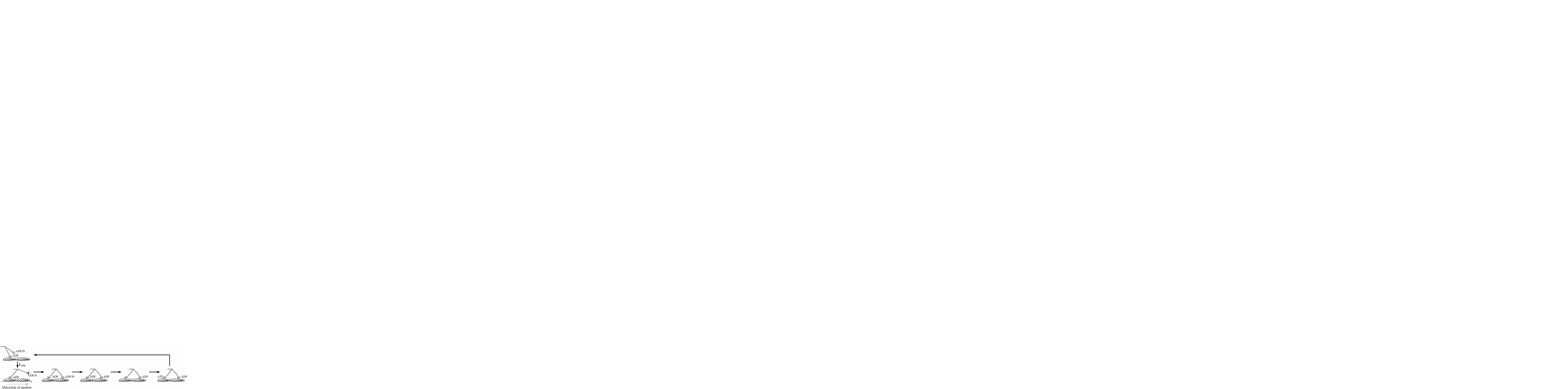}
    \end{center}
    \caption{Duty cycle of a dimeric myosin V motor.  A head with ADP undergoes
      the power-stroke, then the second head containing ADP.Pi binds in the
      lead position, it releases Pi, then the rear head releases ADP, binds a
      new ATP molecule and detaches from actin. After that the cycle repeats
      with exchanged roles of the two heads and with the motor one step
      further.}
    \label{fig:dutycycle}
  \end{figure*}

  \begin{itemize}
  \item The mean run length over a wide range of nucleotide concentrations, but
    without an applied load, has been measured by Baker \emph{et al.}
    \cite{Baker.Warshaw2004}.  It shows a slight decrease with a growing ATP
    concentration.  Added ADP decreases the run length, but only by about a
    factor of 2, which is reached at concentrations around $200\,\rm \mu \rm
    M$.  Under higher ADP concentrations, the run length remains constant at
    around 400nm.  Added $40\,\rm mM$ phosphate led to a reduced run length
    \cite{Baker.Warshaw2004} at zero load and to a lower load at which the
    molecule dissociates \cite{Kad.Warshaw2008}. 

  \item Knowledge on the load-dependence of the run length is still rather
    incomplete. Clemen \emph{et al.}\ \cite{Clemen.Rief2005} measure a largely
    constant run length of about 300nm under loads ranging between -5pN and
    1.5pN.  Under a super-stall load of 5pN, the average backward run length
    was 80nm.

  \item The run length decreases with an increasing ionic strength
    \cite{Baker.Warshaw2004,Hodges.Trybus2007}.  It could be increased by
    adding (through mutation) additional positive charge to loop 2, which
    increases the electrostatic attraction between the myosin V head and actin
    \cite{Hodges.Trybus2007}.
  \end{itemize}
\item These mechanical data are directly complemented by \emph{kinetic
    measurements}, which follow the transitions that take place when motor
  molecules with a certain nucleotide (ATP or ADP) are mixed with actin
  filaments and a different nucleotide in solution.  There are generally two
  ways of probing the myosin V kinetics - using single-headed (usually S1)
  molecules, or double-headed motors.  Experiments on single-headed motors
  reveal the kinetics of the unstrained ATPase cycle.  Experiments with
  double-headed motors, on the other hand, bring additional insight into the
  kinetics in states where both heads are bound and some transitions are
  suppressed (or possibly in some cases accelerated) by intra-molecular strain.

  \textbf{Single-headed:}
  \nopagebreak

  Kinetic studies on single-headed myosin V molecules were pioneered by De La
  Cruz and coworkers.  In Refs.
  \cite{De_La_Cruz.Sweeney1999,Yengo.Sweeney2002}, a number of kinetic rates
  have been determined.  These include the ATP binding rate (both in free heads
  and on actin), the ATP hydrolysis rate, the phosphate release rate on actin,
  the ADP release and re-binding rate (in free/bound heads), and the
  actin-myosin binding/unbinding rate in apo (state without any nucleotide) and
  ADP states.  Trybus \emph{et al.}\  \cite{Trybus.Freyzon1999} report similar
  values, but with a large discrepancy with regard to the ADP release rate in
  free heads.  Wang \emph{et al.}\  \cite{Wang.Sellers2000}, on the other hand,
  report a slower phosphate release rate.  Forgacs \emph{et al.}\
  \cite{Forgacs.White2006} measured the kinetics with deac-aminoATP{}.  In
  summary, one can say that the kinetics of single-headed myosin V is well
  characterised, except for some short-lived states (such as the dissociation
  rate in the ATP state) or rare transitions (such as dissociation in the
  ADP.Pi state).

  \textbf{Double-headed:} 
  \nopagebreak

  Kinetic measurements on double-headed myosin V molecules in the presence of
  actin are crucial for a quantitative understanding processive motility and
  the coordination of chemical cycles.  At the same time, they are much more
  difficult to interpret, because one measures a superposition of contributions
  from the lead head, from the trail head and from molecules bound with a
  single head only.

  The first kinetic study on dimers was carried out by Rosenfeld and Sweeney
  \cite{Rosenfeld.Sweeney2004}.  They found that myosin V with ADP, mixed with
  actin, binds with both heads and releases ADP from the trail head at a rate
  of $28-30\,\rm s^{-1}$ and from the lead head at about $0.3-0.4\,\rm s^{-1}$.
  Note that the ADP release rate from the trail head is somewhat (by about a
  factor of 2) accelerated in comparison with single-headed myosin V, while it
  is strongly suppressed on the lead head.  This is attributed to the effect of
  intra-molecular strain, which affects the ATP release both from the lead and
  from the trail head.  If myosin V with ADP and Pi on it is mixed with actin,
  it releases phosphate from both heads at a measured rate of about $200\,\rm
  s^{-1}$.  However, obtaining this rate involved some extrapolation, so the
  lower values obtained in later studies
  \cite{Yengo.Sweeney2004,Forgacs.White2008} might be more reliable.  But the
  important message of the paper is that the lead head releases phosphate
  quickly and then waits in the ADP state.  ADP release from the lead head was
  therefore identified as the key gating step in the chemical cycle.

  As a result of these kinetic studies, the duty cycle shown in
  Fig.~\ref{fig:dutycycle} has emerged with a high level of consensus.

\end{enumerate}

\section{Open questions}
\label{sec:open-questions}

Yet some of the important questions on myosin V have only been partially
answered so far.  In the following, we will discuss five major open questions.
They are, on one side, important for setting up theoretical models.  At the
same time, most of these questions are quantitative in nature and most of them
concern properties that are not directly observable in an experiment.
Therefore, they can only be tackled efficiently by combining experimental
results with theoretical models.

\subsection{Head-head coordination}

In order to maintain its processivity, a dimeric motor has to prevent
simultaneous detachment of both heads from its track. This means that the
chemical cycles of both heads have to be coordinated in order to stay out of
phase. There is a high level of agreement that this coordination is achieved
through intra-molecular strain.  There is also undisputed evidence for
\emph{lead head gating}, which means that ADP release from the lead head waits
until the trail head detaches from actin
\cite{Rosenfeld.Sweeney2004,Purcell.Spudich2005,Forgacs.White2008}.  However,
there is also some evidence for much weaker gating in the trail head, meaning
that the ADP release rate in the trail head is somewhat accelerated by the
presence of the bound lead head \cite{Veigel.Molloy2002,Rosenfeld.Sweeney2004}.

Both effects are likely connected with the small (about $5\,\rm nm$ for
single-headed molecules) conformational change that takes place upon ADP
release \cite{Veigel.Molloy2002,Veigel.Sellers2005}. Conversely, Oguchi
\emph{et al.}\ \cite{Oguchi.Ishiwata2008} have recently directly measured the
influence of an applied load on the ADP binding affinity. The strong
deceleration of ADP release in the lead head is possibly also connected with
the large main power stroke.  The exact nature of this influence depends on
whether the ``telemark state'' exists or not (see next section).

In addition, coordination can take place during other steps, too.  The elastic
lever arm model shows that the attachment rate of the lead head is strongly
suppressed before the trail head goes through a power-stroke \cite{Vilfan2005}.

  \begin{figure}
    \begin{center}
      \includegraphics{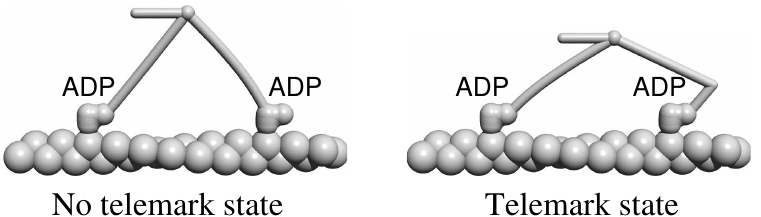}
    \end{center}
    \caption{In the model without a telemark state (left), the lead head stays
      in the pre-powerstroke state before the trail head detaches.  In the
      telemark state (right), both heads are in the post-powerstroke state, but
      the leading lever arm has a strong kink -- resembling a telemark skier.}
    \label{fig:telemark}
  \end{figure}

\subsection{Lead head state} 

One of the most controversial questions on myosin V is whether the lead head is
in the pre-powerstroke state, or in the post-powerstroke state with a strongly
bent lever arm (also called ``telemark'' state, because the shape of the
molecule resembles a telemark skier with the front knee bent), as shown in
Fig.~\ref{fig:telemark}.  Evidence for the telemark state first came from EM
images \cite{Walker.Knight2000}, although subsequent analysis revealed that
molecules with a kinked leading lever arm are not as abundant as those with a
straight one \cite{Burgess.Trinick2002}, and that the telemark state might not
lie on the main pathway.  Further evidence came from experiments with
fluorescence polarisation \cite{Forkey.Goldman2003,Snyder.Selvin2004}, which
showed that about a third of labels attached on the lever arm did not show any
detectable tilting motion. But later studies revealed that this was probably a
consequence of the ambiguity in angle detection from one vector component alone
and that all labels on the lever arm do indeed show tilting motion, as well as
alternating steps \cite{Toprak.Selvin2006,Syed.Goldman2006}.

However, it is still possible that the head domain of the lead head is in some
intermediate state and that its lever arm is bent, but not as strongly as in
the pure telemark state.

  \begin{figure}
    \begin{center}
      \includegraphics{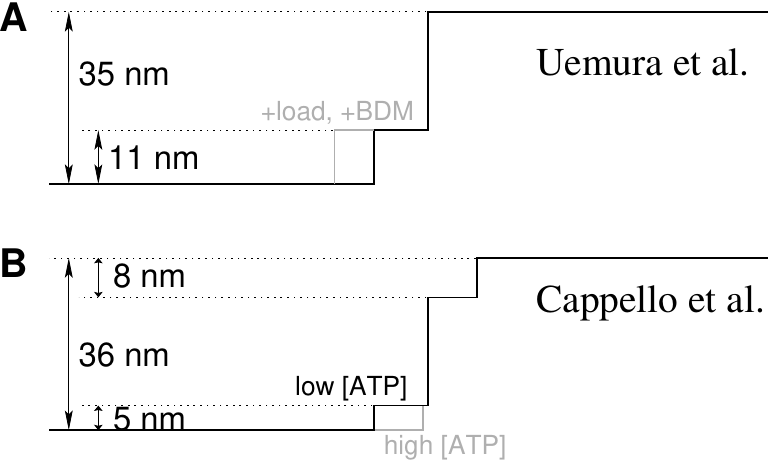}
    \end{center}
    \caption{Sub-steps as observed by Uemura \emph{et al.}\ (A) and Cappello \emph{et al.}
      (B).  Uemura \emph{et al.}\ observe a substep of 11nm, preceding the main step.
      Its size is independent of load, but its duration increases with load or
      upon addition of BDM.  Cappello \emph{et al.}\ observe one substep preceding the
      main step (only at low ATP concentrations) and one following the main
      step (independent of ATP concentration).}
    \label{fig:substeps}
  \end{figure}

\subsection{Sub-steps}

An intriguing question is whether the motor moves its load in uniform steps,
or do they contain short-lived substeps.  Because of the noise in position
determination, these sub-steps are difficult to detect.  The experiment of
Uemura and co-workers \cite{Uemura.Ishiwata2004} revealed substeps of 11+24nm,
as shown in Fig.~\ref{fig:substeps}A.  Their length was independent of load,
but their duration was increased both with an increasing load and upon
addition of BDM (butanedione monoxide). The duration was independent of the
ADP and ATP concentration.

The second important observation related to sub-steps comes from Cappello
\emph{et al.}\ \cite{Cappello.Prost2007}. This study employs travelling wave
tracking, a technique that allows a significantly higher temporal resolution.
It reveals two substeps of about 5nm: one immediately preceding and one
immediately following the main step (Fig.~\ref{fig:substeps}B). The duration of
the substep before the main step was longer at a lower ATP concentration.
Therefore, neither of the two sub-steps has a direct correspondence in the
observation of Uemura \emph{et al.} The substep following the main step can
well be attributed to the diffusive search of the lead head for the next actin
binding site.  The explanation for the step preceding the main step is more
difficult.  The authors propose a hypothesis in which the substep represents a
reversible transition in the lead head that can take place after the trail head
releases ADP and which is a necessary condition for ATP binding to the trail
head.  One question that remains open in this scenario is the mechanism by
which ATP binding to the trail head could be triggered by a transition in the
lead head.

Sub-steps have also been observed in two somewhat different experiments.
First, Dunn and Spudich \cite{Dunn.Spudich2007} have labelled sites on the
lever-arm with gold nanoparticles and tracked it with sub-millisecond time
resolution.  They found that the $74\,\rm nm$ steps were subdivided into
$49\,\rm nm + 25\,\rm nm$ and that the intermediate state had a very high level
of fluctuations, resulting from the free head diffusively searching for the
next binding site.  Depending on the light chain presence, the lead head
binding rate was either $60\,\rm s^{-1}$ or $180\,\rm s^{-1}$.  It was slowed
down by the presence of BDM -- different from the conclusion of Uemura \emph{et
  al.}\ \cite{Uemura.Ishiwata2004} that the effect of BDM is a delayed power
stroke.  The lead head binding rate is somewhat slower (possibly because of the
attached particle), but still of the same order of magnitude as the duration of
the second sub-step ($1-2\,\rm ms$) measured by Cappello \emph{et al.}\
\cite{Cappello.Prost2007}.  A direct comparison of step sizes between the two
experiments is difficult, because Cappello \emph{et al.}\ tracked the cargo
position, whereas Dunn and Spudich tracked a bead attached to the lever arm.

Finally, as already mentioned, a sub-step of about $5\,\rm nm$ has been
observed by Veigel \emph{et al.}\ \cite{Veigel.Molloy2002,Veigel.Sellers2005}
in single-headed myosin V molecules and identified as the conformational change
upon ADP release.  There are good reasons to believe that this sub-step is not
directly observable in dimeric motors \cite{Vilfan2005}.  One argument is that
ADP release shows a strong load dependence in single-headed motors
\cite{Veigel.Sellers2005,Oguchi.Ishiwata2008}.  In dimeric motors, moderate
backward loads up to $\sim 1\,\rm pN$ or forward loads show almost no effect on
the stepping rate, which is (under saturating ADP concentrations) limited by
ADP release \cite{Mehta.Cheney1999,Uemura.Ishiwata2004}.  However, Cappello
\emph{et al.}\ hypothesise that the conformational change upon ADP release in
the trail head might be involved in triggering a reversible transition in the
lead head, which then causes an observable sub-step \cite{Cappello.Prost2007}.

\subsection{How tight is the coupling between the hydrolysis cycle and the
  stepping?}

The Michaelis-Menten like dependence of the kinesin velocity on the ATP
concentration \cite{schnitzer97} was seen as early evidence for a tight
coupling between ATP hydrolysis and stepping (in fact, those experiments could
not yet exclude additional futile ATP hydrolysis cycles).  Similar experiments
also revealed tight coupling between stepping and ATP hydrolysis in myosin V
under low loads \cite{Rock.Spudich2000}.  Therefore, early kinetic models
assumed ATPase cycles tightly coupled to the stepping
\cite{Fisher.Kolomeisky2001,Kolomeisky.Fisher2003}, which means that the motor
always makes one step when it hydrolyses one ATP molecule and that it can only
make a backward step upon re-synthesis of an ATP molecule from ADP and
phosphate.
 
Such reversible behaviour has indeed been demonstrated in the $\rm F_1$-ATPase,
which can either hydrolyse or synthesise ATP, depending on the direction of
rotation \cite{Itoh.Kinosita2004}. However, experiments on both myosin V and
kinesin have shown that tight coupling has its limits under high loads.  For
kinesin, Carter and Cross \cite{Carter.Cross2005} found out that under high
loads the motor is pulled backwards, for which it requires ATP binding.  The
observation that myosin V can be pulled backwards by a super-stall force was
first made by Clemen and coworkers \cite{Clemen.Rief2005}.  The backward and
forward steps were further investigated in a subsequent paper
\cite{Gebhardt.Rief2006}, where they found out that forward steps steps involve
ATP hydrolysis, while backward steps involve neither hydrolysis nor synthesis
of ATP.  A rather surprising consequence of this finding is that a high load
(e.g., 5pN) can pull the motor faster when applied in the backward direction
than in forward.

So under sufficiently high loads, both myosin V and kinesin leave the tightly
coupled cycle, but each in a different way: kinesin can be pulled backwards in
the presence of ATP, whereas myosin V neither hydrolyses nor synthesises it.

\subsection{Source of elasticity}

In any mechano-chemical model for myosin V, there needs to be at least one
elastic element, such as the lever arm.  One reason is that some degree of
flexibility is necessary in order for the motor to be able to bind to the actin
filament with both heads simultaneously.  Furthermore, many models propose that
when a head commits its power stroke, some of the free energy released is first
stored into elastic energy and later uses it to perform work on the load.
Besides the lever arm, which could be an obvious elastic element, further
candidates for the source of compliance lie in or around the converter domain.
In particular, the converter domain was shown to be the main source of
compliance in myosin II \cite{Kohler.Kraft2002}.  However, the same study
showed that the converter domain was significantly stiffened through a single
mutation, which means that the result is not directly transferable between
different myosin classes.
  
Measuring the stiffness directly requires careful elimination of other sources
of compliance, such as the actin filament and the tail domain. Veigel \emph{et
  al.}\ measured the stiffness of single-headed myosin V and obtained the value
0.2pN/nm \cite{Veigel.Sellers2005}.  A comparison with smooth muscle myosin led
to the suggestion that the elastic element is located close to the pivot point
of the lever arm.  Of course, this conclusion cannot be regarded as definite,
because the degree of mechanical similarity between the two types of myosin is
not known.

On the other hand, the existence of a telemark state would suggest that the
lever arm has a very compliant point (sufficiently compliant that it can be
bent by the power-stroke), possibly around the 4th IQ domain.  MD simulations
\cite{Ganoth.Gutman2007} support the possibility of such a kink, but are not
yet capable to precisely predict the bending energy connected with it.

\section{Theoretical models}
\label{sec:theoretical-models}

\subsection{Overview}

\begin{table*}[!t]
  \caption{Comparison of five theoretical (kinetic and mechano-chemical)
    models. ``$+$'' denotes that the given experimental observation is in agreement
    with the model.  ``$\circ$'' means partial or qualitative agreement, ``$-$'' denotes
    disagreement, ``n/a'' (not applicable) that the given aspect is not
    included in the model or that no analysis is available that would allow a
    comparison.}
  \label{tab:model-comparison}
  \begin{minipage}{\textwidth}
\renewcommand{\thempfootnote}{\fnsymbol{mpfootnote}}
  \begin{tabular}{|p{5.5cm}|c|c|c|c|c|}
    \hline
    & Kolomeisky \cite{Kolomeisky.Fisher2003} & Skau \cite{Skau.Turner2006} & Wu
    \cite{Wu.Karplus2007} & Lan \cite{Lan.Sun2005} & Vilfan (5-st.) \cite{Vilfan2005}  \\ \hline
    \textbf{Force-velocity relation}:&&&&&\\ \hline
    Flat region around zero force & $\circ$ &  $\circ$ & n/a & $+$     & $+$    \\ \hline
    Flat for high forward loads & $-$       &  $+$     & n/a & $\circ$ & $+$    \\ \hline
    Around stall                & $+$       &  $+$     & n/a & $+$     & $+$    \\ \hline
    Above stall                 & $\circ$\footnote{Model shows backward stepping, but
      only through ATP synthesis}       &  $-$     & n/a &
    $\circ$\footnote{Velocity too high; dependence on ATP concentration not
      available}\footnote{In the refined model \cite{Lan.Sun2006} no
      backward stepping takes place above 3pN load.} & $\circ$\footnote{Velocity too low} \\ \hline
    [ATP] dependence            & $+$       &  $+$     &  $+$     &  n/a     &  $+$
    \\ \hline  
    [ADP] dependence            & n/a       &  $+$     &  $+$     &  n/a     &  $+$
    \\ \hline  
    \textbf{Substeps}:&&&&&\\ \hline
    As observed by Uemura \cite{Uemura.Ishiwata2004}\newline (11nm+24nm, load independent) & $+$ & $-$ & $+$ 
    & & $-$ \\ \cline{1-4} \cline{6-6}
    As observed by Cappello \cite{Cappello.Prost2007},\newline before main step (only at low [ATP])&  $-$  & $-$  & $-$\footnote{Model shows a substep upon trail head release, but predicts an
      incidence that increases with the ATP concentration.}& ?\footnote{Histograms of position changes show
      intermediate peaks, but no analysis is available which would attribute them to any type of sub-steps.} & $-$\footnote{Model shows a substep upon trail head
      release, but it does not
      depend on ATP concentration}  \\  \cline{1-4} \cline{6-6}
    As observed by Cappello  \cite{Cappello.Prost2007},\newline after main step&  $-$  & $+$  & $-$ &  & $+$ \\ \hline
    \textbf{Processivity (run length)}:&&&&&\\ \hline
    [ATP] dependence at zero load & n/a & $+$ & $\circ$ & n/a & $+$ \\ \hline
    [ADP] dependence at zero load & n/a & $-$ & $+$     & n/a & $\circ$ \\ \hline
    Load dependence: super-stall forces & n/a & $-$ & n/a & n/a & $+$ \\ \hline
    Load dependence: forward forces & n/a & $+$ & n/a & n/a & $\circ$\footnote{Model processive only up to  3pN forward load}\\ \hline
  \end{tabular}
\end{minipage}
\end{table*}

Theoretical models can be employed to describe the dynamics of a motor
protein on different levels.  First, there are \textit{discrete stochastic
  models} that describe the motor with a collection of chemical states that
correspond to different positions along the track \cite{Kolomeisky.Fisher2007}.
The motor protein is treated as a ``black box'', the only observable being the
tail- or the cargo position. This approach is particularly useful for
interpreting experimental results gained with optical tweezers, such as
force-velocity relations under different ATP, ADP and phosphate concentrations.
It was first applied to myosin V by Kolomeisky and Fisher
\cite{Kolomeisky.Fisher2003}. The basic version of a discrete stochastic model
allows no events beyond the regular chemical cycle, but it can be extended to
allow parallel pathways
\cite{Skau.Turner2006,Wu.Karplus2007,Tsygankov.Fisher2008}, as well as
dissociation from the actin filament.  Coordinated hand-over-hand motility is
one of the assumptions of such models, rather than being its outcome.

The next level of models uses a similar phenomenological approach to describe
each individual head.  The properties of a dimeric motor are then derived from
those of the two heads.  This approach is similar to that taken by muscle
myosin models \cite{huxley57,hill74,vilfan2003b}, which deal with a large
number of myosin heads, assembled into thick filaments.  Such models combine
the head domain, described with a few discrete chemical states, with an elastic
element that connects it to the other head and to the tail.  We will therefore
use the term \emph{mechano-chemical models} for this category.  Two models for
myosin V that follow this approach are described in Refs.  \cite{Vilfan2005}
and \cite{Lan.Sun2005}.  In theory, the mechano-chemical models can be mapped
onto corresponding discrete stochastic models.  A state in the resulting
discrete stochastic model is then described with a combination of the lead head
state, the trail head state and the relative distance of the two heads on the
track.  Without further simplifications, the number of such states is therefore
usually too high to be studied analytically.

Finally, there are molecular models for motor proteins that attempt to
understand the mechanism of energy transduction in microscopic detail.  A full
scale molecular dynamics simulation over millisecond time scales on which the
chemical cycle takes place is still well beyond the available computational
power, but it can be used to probe essential processes in the course of ATP
hydrolysis \cite{Schwarzl.Fischer2006,Koppole.Fischer2006,Yu.Cui2007}. Also,
several approximative methods have been employed.  The dynamics of the power
stroke has been investigated with molecular kinematics using the minimum energy
path method (MEP) \cite{Fischer.Smith2005,Koppole.Fischer2007}. While these
computational models represent an important contribution to our understanding
of the microscopic mechanism by which molecular motors work, they are currently
not yet precise enough to make predictions on free energy differences between
different states or transition rates between them.  Therefore, quantitative
models still need to be ``reverse-engineered'' to a large extent from
experimental data.

There are other noteworthy models that deal with special aspect of myosin V,
but will not be the subject of our review.  Here we should briefly mention the
work of Smith, which relates the duty ratio of myosin V to its processivity
\cite{Smith2004}.  Further, there is a study by Schilstra and Martin
\cite{Schilstra.Martin2006} who investigate how viscous load can decrease the
randomness of the walking molecule.  Finally, a number of models deal with the
collective dynamics of motors in general \cite{vilfan98,Klumpp.Lipowsky2005},
and those findings could be relevant for the dynamics of myosin V transporting
vesicles, too.

In the following, we will give an overview over 3 discrete stochastic models
and 2 mechanochemical ones.  In particular, we will identify the common points
and the differences between them.  Table \ref{tab:model-comparison} shows an
overview which model properties are in agreement with which experimental
results and where the discrepancies are.

\begin{figure*}
\begin{center}
\begin{tabular}{lr}
\multicolumn{2}{l}{\raisebox{4cm}{\textsf{\textbf{A}}}
  \includegraphics{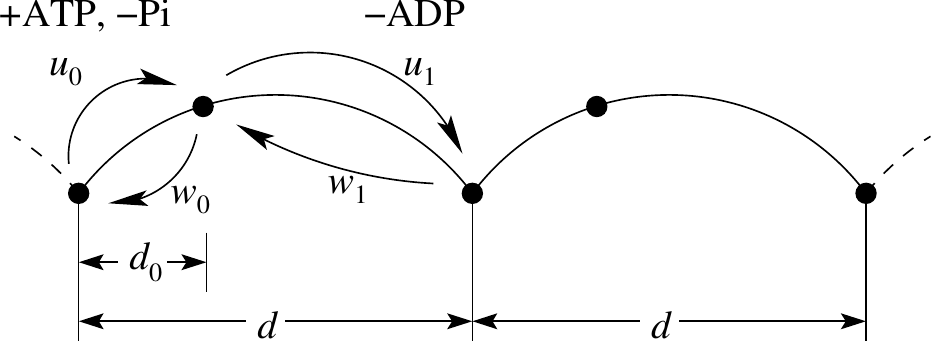}} \\ \hline\\
\raisebox{4cm}{\textsf{\textbf{B}}} \includegraphics{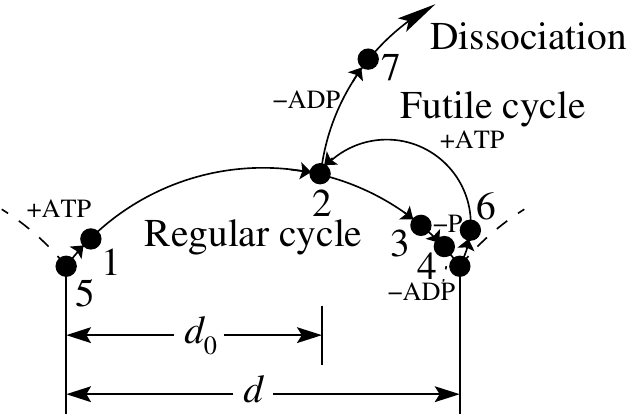} &
\includegraphics{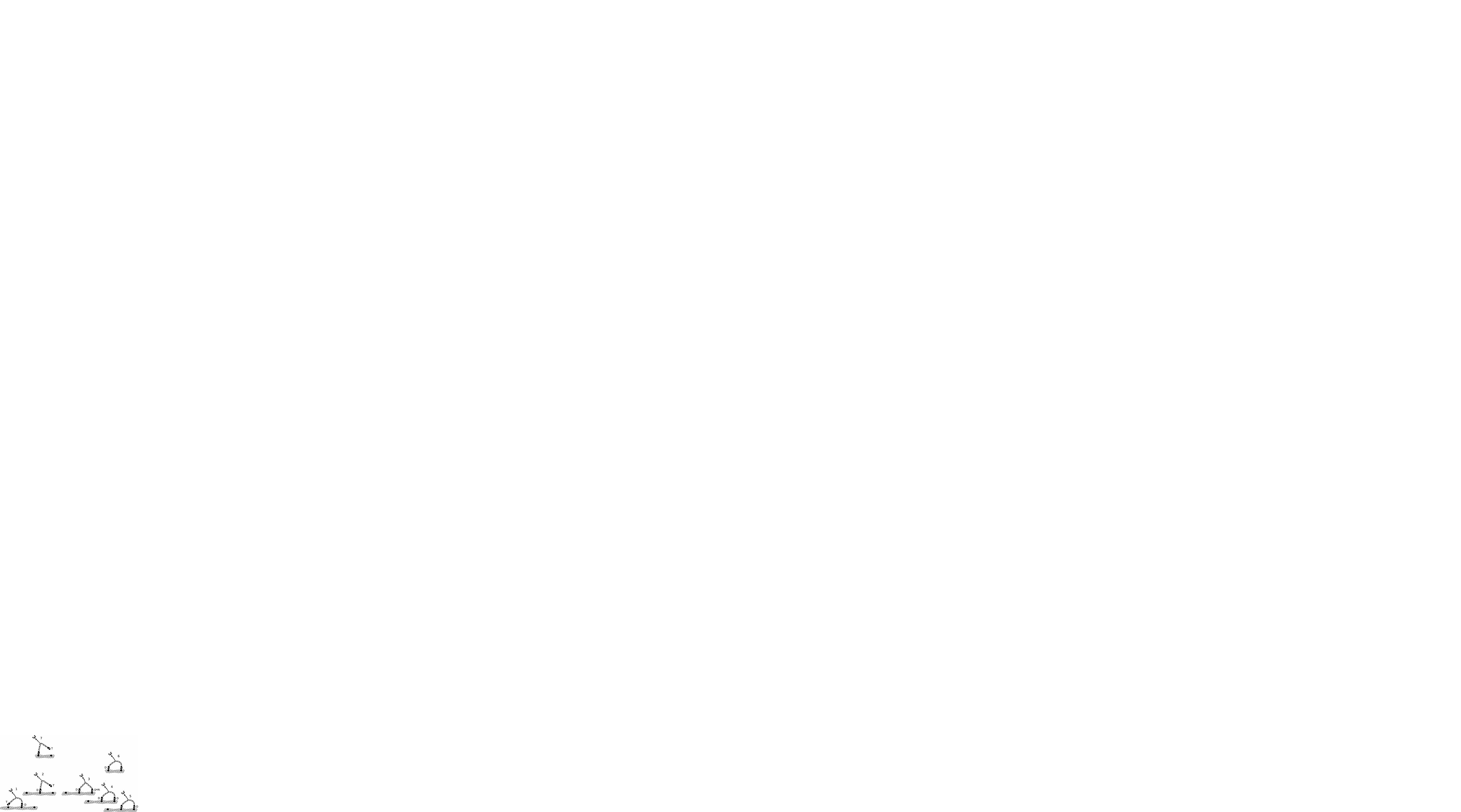} \\ \hline\\
\raisebox{4cm}{\textsf{\textbf{C}}}  \includegraphics{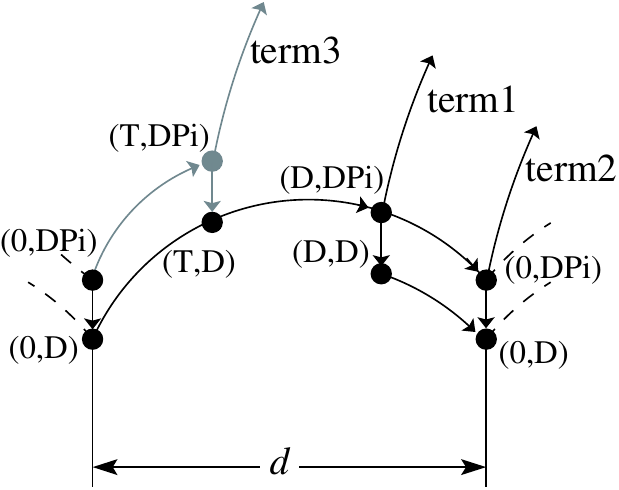}&
\raisebox{4cm}{\textsf{\textbf{D}}}  \includegraphics{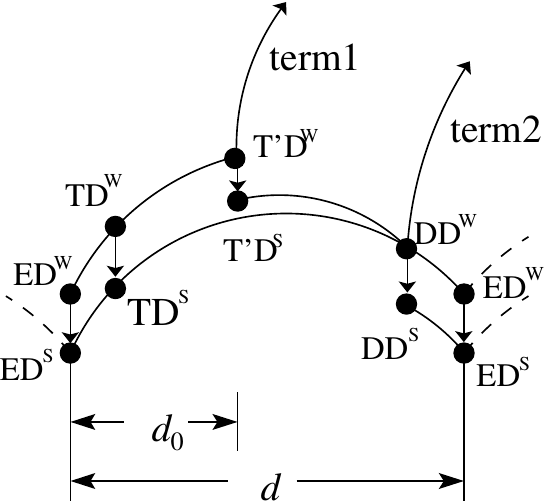}
\end{tabular}
\end{center}
\caption{Four kinetic models with different levels of complexity. (A) Model by
  Kolomeisky and Fisher \cite{Kolomeisky.Fisher2003} (B) Model by Skau, Hoyle
  and Turner \cite{Skau.Turner2006}. The right part shows the states numbered
  1-7. Note that states 1,3,4,5 and 6 have the same load position.  (C) Model
  by Baker et al.\ \cite{Baker.Warshaw2004} (D) Model by Wu, Gao and Karplus
  \cite{Wu.Karplus2007}.}
  \label{fig:kineticschemes}
\end{figure*}

\subsection{Discrete stochastic models}
\subsubsection{Kolomeisky and Fisher} 

Kolomeisky and Fisher \cite{Kolomeisky.Fisher2003}
(Fig.~\ref{fig:kineticschemes}A) presented the first theoretical model
describing myosin V.  It uses 2 states per step.  The intermediate state
corresponds to a substep of about 13nm, measured from the ATP binding state.
The model parameters were determined by fitting the calculated force-velocity
relations to data from Ref.~\cite{Mehta.Cheney1999}.  The conclusions of the
study are that both the ATP binding rate is load-independent and the ADP
release rate is only weakly load dependent -- which implies that most of the
load dependence is contained in their reverse rates.  At super-stall forces,
the model shows backward steps, but they require ATP synthesis. Besides fitting
the load velocity relations, the model also makes predictions on the randomness
parameter of the trajectory \cite{svoboda94b}, defined as
\begin{equation}
  \label{eq:randomness}
  r=\lim_{t\to \infty}\frac {\left<x^2(t)\right>-\left< x(t) \right>^2}{a \left< x(t) \right> }\;,
\end{equation}
where $a$ denotes the step size and $\left< \ldots \right>$ an ensemble
average. The value of $r$ is 1 for a stepper with a single rate limiting step.
A higher value of indicates the presence of backward- or multiple steps, while
$r<1$ means that several rate limiting transitions per step are involved.  For
myosin V, the randomness parameter has not yet been measured to date.

Prediction of substeps of a similar size as later observed by Uemura \emph{et
  al.}  \cite{Uemura.Ishiwata2004} is a particular merit of this model.
However, it should be mentioned that the observed substeps had a duration
around 10ms.  In the model, the duration of substeps corresponds to the ADP
release time, which is around 80ms.  This discrepancy is an indication that two
states per step are not sufficient if one wants to describe the myosin V motion
in this amount of detail.

\subsubsection{Skau, Hoyle and Turner} 

Skau \emph{et al.}\ \cite{Skau.Turner2006} (Fig.~\ref{fig:kineticschemes}B)
developed a more elaborated myosin V model, which still belongs to the category
of discrete stochastic models, but includes two parallel cycles - one that
leads to a forward step and a futile one that does not.  In addition, one
dissociation path from actin is included.  The model does not include backward
stepping without ATP synthesis.  Substeps of 25+11nm are included in the model,
but in reverse order compared with the model by Kolomeisky and Fisher
\cite{Kolomeisky.Fisher2003} and the experimental observation by Uemura et al.\
\cite{Uemura.Ishiwata2004}, where ATP binding is followed first by a short
(12nm) and then the longer (24nm) substep, rather than the long step followed
by the short one.

The model fits well the force-velocity relations, although the velocity starts
to drop around zero force already, rather than being nearly flat up to about
1pN, as experimental data suggest.  Because the model does not include backward
stepping by other means than reversing the ATP hydrolysis, it fails to
reproduce the velocity at super-stall loads.

The calculated run length increases inversely proportional with a decreasing
ATP concentration.  It also increases with an increased ADP concentration --
the latter is inconsistent with the measurement of Baker et al.\
\cite{Baker.Warshaw2004}.

\subsubsection{Wu, Gao and Karplus} 

Wu \emph{et al.} \cite{Wu.Karplus2007} (Fig.~\ref{fig:kineticschemes}D)
recently proposed a model that contains more kinetic details, but no external
force (load).  In that model, each head can be found in the empty state, in the
ATP state, in the detached ATP state, in the weakly bound ADP state and in the
strongly bound ATP state.  The trail head can either be in the empty, ATP, or
strongly bound ADP state.  The lead head can be in the strongly or weakly bound
ADP state.  The chemical cycle can follow three different pathways (mainly
depending on when the weakly- to strongly-bound transition in the leading head
takes place).  Upon trail head detachment, the tail moves by 12nm,
corresponding to the substep from Ref.  \cite{Uemura.Ishiwata2004}.  This
substep occurs only on one of the three kinetic pathways, which is in agreement
with the experiment that showed substeps only for a fraction of steps. There
are also two different termination pathways, which lead to the dissociation of
the whole dimer from actin.

The main results of the model are zero-load velocity and run length as a
function of ATP and ADP concentration.  The calculated run length shows a
decrease with the ADP concentration, which is in good qualitative agreement
with Ref.\ \cite{Baker.Warshaw2004}.  The run length increases somewhat with an
increasing ATP concentration, which is different from the conclusion of Baker
and coworkers \cite{Baker.Warshaw2004}, although, given the uncertainty of data
points, not necessarily inconsistent with their measurements.  The model shows
that the velocity at low ATP concentrations is very sensitive to ADP, which
could possibly provide a reconciliation of the large differences between
different experimental results.

A question that is beyond the scope of a purely kinetic model, but will have to
be answered when mechanical aspects are included, is how the intra-molecular
strain can have such a strong influence on the weakly- to strongly-bound
transition in the lead head.  In the model, the transition rate decreases
almost 4-fold when the trail head releases ADP.  Another puzzle brought up by
the model is why the dissociation rate is 30 times higher in a state where both
heads are bound to actin (one weakly and one strongly) than in a state where
only one head is weakly bound to actin and the other one is free.  While this
seems inevitable for an explanation of run length data by Baker \emph{et al.}
\cite{Baker.Warshaw2004}, it is not easy to think of a mechanical model in
which myosin V, bound with both heads, is more likely to dissociate than when
bound with a single head.  Internal strain might potentially provide an
explanation, but only if the dissociation takes place in a single step and not
with one head after the other.

\subsection{Mechano-chemical models}

The main difference between the models listed above and the mechano-chemical
approach is that the latter uses a \emph{single myosin V head} as the model
unit and derives the properties of a dimer.  This reduces the number of
parameters and model assumptions.  The model is specified by the states of an
individual head (typically 4 or 5 bound states, plus a number of free states),
by the mechanical conformation of the molecule in that state, by the transition
kinetics and by the interaction between the two heads that is mediated by the
mechanical properties of the lever arms joining them.  In practice, however,
even the kinetics of a single head contains a number rates that have not yet
been measured directly and that need to be determined from dimer properties,
such as force-velocity relation and processivity.  So far, two myosin V models
have been developed on these premises: the model by Lan and Sun
\cite{Lan.Sun2005} and the elastic lever arm model \cite{Vilfan2005}.  Although
they share the basic principle, there are significant differences in the
implementation.

\subsubsection{Lan and Sun} 
Lan and Sun \cite{Lan.Sun2005} proposed a model based on a cycle with 5
chemical states, all of which can in theory be bound to actin or free.  The
transition rates are taken from kinetic studies or guessed on the basis of
other myosins where no data are available.  Nonetheless, some assumed rate
constants differ significantly from generally accepted values.  For instance,
the detachment rate of a head from actin in the empty state is assumed to be
$0.16\,\rm s^{-1}$, while the value from Ref.\ \cite{De_La_Cruz.Sweeney1999} is
$0.00036\,\rm s^{-1}$.
  
The elasticity of the two lever arms is modelled as highly anisotropic.  It
combines an in-plane component and a very stiff azimuthal component. The
in-plane elasticity corresponds to a persistence length of 120nm, or a spring
constant of $0.05\,\rm pN/nm$, measured at the tip.  This is consistent with an
earlier estimate based on tropomyosin \cite{Howard.Spudich1996}, but much lower
than the stiffness measured with optical tweezers ($0.2\,\rm pN/nm$
\cite{Veigel.Sellers2005}).  The azimuthal component is modelled in an \emph{ad
  hoc} way and is assumed to be very stiff.  If the model molecule binds to two
adjacent sites on the actin filament, the bending energy is $40\,k_BT$,
compared to $15\,k_BT$ in the model with isotropic lever arm elasticity
\cite{Vilfan2005}.

In contrast with the model in Ref.\ \cite{Vilfan2005}, the heads also possess
internal flexibility, meaning that the lever arm angle can fluctuate to some
extent while the head stays in the same chemical state.  This allows the
transitions to take place without the need for a simultaneous lever arm swing.

The main result of the model is the force-velocity relation for a fixed set of
nucleotide concentrations.  It reproduces well the flat region at low loads and
the abrupt drop around 1pN, but there is a discrepancy above the stall force,
where it predicts a backward speed well above experimental values.  No
processivity calculations are made in the paper, so it is likely that the
parameters will need adjustment to achieve realistic run lengths.

A subsequent paper by the same authors \cite{Lan.Sun2006} deals primarily with
myosin VI, but it also presents a refined model for myosin V.  This includes a
simpler kinetic model, described with a 4-state cycle. The second important
change is that the lever arm elasticity is modelled in a more realistic way, as
an anisotropic elastic beam with persistence length values of $150\,\rm nm$ and
$400\,{\rm nm}$ in two different planes.  The elastic energy in a state where
the two heads are separated by two actin subunits is reduced to about $10\,k_B
T$, as compared with $40\,k_B T$ in the original model.  The calculated
force-velocity relation remains similar as in the older model for forward and
below-stall loads.  For super-stall forces, the new model predicts nearly
negligible backward velocities.

\subsubsection{Elastic lever arm model} 

\begin{figure}\hspace*{-.5cm}
\includegraphics[scale=0.9]{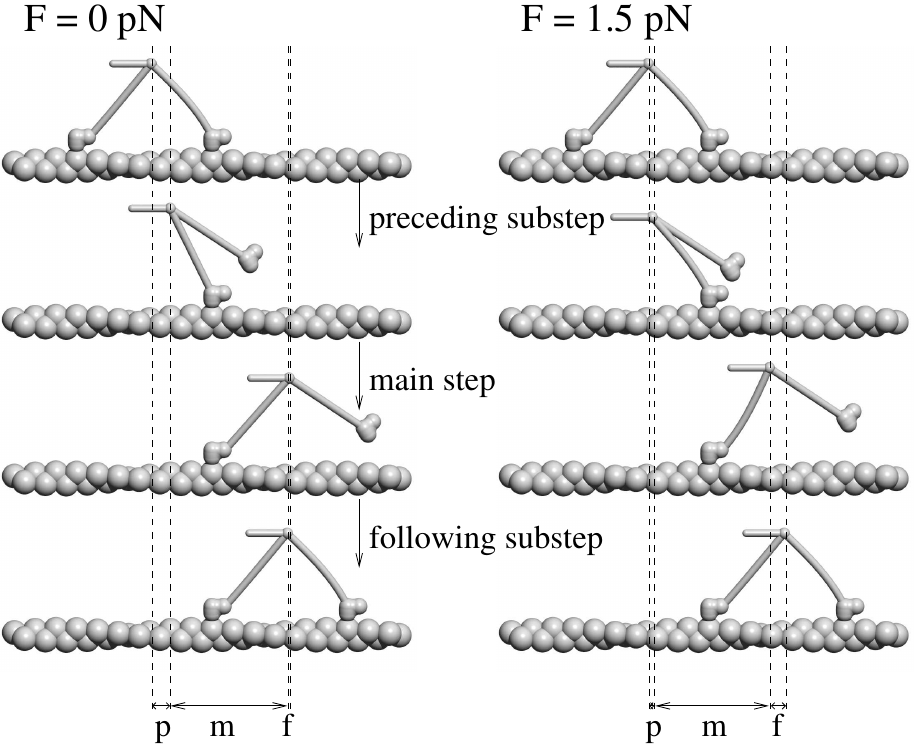}
  \caption{Substeps in the elastic lever arm model.  Because they involve changes between configurations bound with a single and with both heads, their size depends on load.}
  \label{fig:substeps-series}
\end{figure}

The elastic lever arm model \cite{Vilfan2005} arose from a similar concept, but
has many differences in the implementation.  The paper presents two alternative
chemical cycles: the 4-state and the 5-state cycle.  The 5-state cycle allows
phosphate release from the lead head before the power-stroke, while the 4-state
cycle does not.  Note that model calculations and also later experimental
evidence strongly favour the 5-state scenario, so we will focus our discussion
of the model on it.  The model uses a cycle of 4 attached and one detached
state (plus the detached ADP state, which is not part of the regular cycle).
Some short lived states (e.g., bound with ATP) are omitted, as are some
unprobable ones (e.g., unbound head in the empty state).

The lever arm is modelled as an isotropic elastic beam with a bending modulus
of $1500\,\rm pN/nm$, which corresponds to a persistence length of $360\,{\rm
  nm}$, or a spring constant of $0.25\,{\rm pN/nm}$ measured at the lever arm
tip.  Both lever arms are connected with a fully flexible joint (the validity
of this assumption was recently proven by Dunn and Spudich
\cite{Dunn.Spudich2007}).

Based on the above value for the elasticity of the lever arm we estimated that
the lead head cannot commit its power-stroke before the trail head detaches,
meaning that the telemark state does not occur.  However, we cannot fully
exclude that nonlinearities in the elastic properties of the lever arm still
sometimes allow this configuration.

The fact that ADP release takes place in the trail head while the lead head is
attached to actin provides a direct explanation for the flat force-velocity
curve below 1pN load.  Namely, although the ADP release is connected with a
small conformational change in the trail head, it causes virtually no load
movement.  Conversely, the load therefore has no influence on the ADP release
rate, which in turn determines the velocity.  For higher loads, close to stall,
other transitions (power stroke or lead head binding) become rate limiting and
the velocity then drops abruptly.

The model reproduces well the force-velocity relation up to the stall load.
For super-stall forces, the model predicted backward slippage without ATP
hydrolysis.  It occurs when both heads have ADP on them.  Very similar
behaviour was later discovered by Gebhardt and coworkers
\cite{Gebhardt.Rief2006}.  However, the backward sliding speed in the model is
too low.  Another discrepancy is that experimentalists observe backward
slippage even in the absence of any nucleotide, while the model needs at least
ADP in order to move backwards. This is due to the assumption that the
detachment rate of a head with ADP or with no nucleotide is the same for the
lead and the trail head, and that it is independent of strain.  In order to
quantitatively fit the newer data, a strain-dependent detachment rate in empty
and ADP states will therefore be necessary.

The model shows a short substep preceding the main step, which is connected
with the trail head detachment.  It also shows a sub-step following the main
step, which corresponds to the binding of the lead head.  The second substep is
in good agreement with the observation of Cappello \emph{et al.}\
\cite{Cappello.Prost2007}.  The first substep is not fully compatible with
their experimental result, because it shows no dependence on the ATP
concentration.  In total, one can still say that the substeps from that model
are much closer to the experimental results of Cappello \emph{et al.}
\cite{Cappello.Prost2007} than to those of Uemura \emph{et al.}
\cite{Uemura.Ishiwata2004}.  In both intermediate states before and after the
main steps the motor is bound with a single head and therefore has a much
higher compliance.  Therefore, both substep sizes depend on the load.  The
first substep decreases with load (Fig. \ref{fig:substeps-series}), while the
second increases.  Uemura \emph{et al.}\ \cite{Uemura.Ishiwata2004}, on the
other hand, see load-independent substep sizes.
  
Another major advantage of the mechano-chemical approach is that dissociation
events that limit the run length are an inevitable and integral part of the
model and do not need any additional assumptions or parameters.  However, they
do depend on several of the kinetic constants whose values are not yet well
known to date.  The model shows that there are generally 3 classes of
dissociation pathways.

  \begin{table*}
    \caption{Dissociation pathway ``dictionary''. The first column denotes the
      state of the dimer from which dissociation occurs.  For example
      (A.M.T,A.M.D.Pi) means that the trail head is bound to actin with ATP on
      it, while the lead head is bound to actin with ADP and Pi.  The other
      columns show the names of these dissociation pathways in different
      models.  Evidence exists that ``Path 3'' is the predominant dissociation
      path.}
    \label{tab:termination-dictionary}
\begin{minipage}{\textwidth}
\renewcommand{\thempfootnote}{\fnsymbol{mpfootnote}}
    \begin{tabular}{|c|c|c|c|c|}\hline
      \parbox{3.5cm}{Pre-dissociation state\\(trail,lead)}& Vilfan
      \cite{Vilfan2005} & Baker \emph{et al.}\ \cite{Baker.Warshaw2004} & Wu \emph{et al.}\ \cite{Wu.Karplus2007} &
      Skau \emph{et al.}\ \cite{Skau.Turner2006}\\
      \hline
      (A.M.T, A.M.D.Pi\footnote{weakly bound A.M.D in \cite{Wu.Karplus2007}})& Path 1 & \multirow{2}{*}{ \parbox{2cm}{term3\\(excluded)}} & term1 & - \\ \cline{1-2} \cline{4-5}
      (A.M.T, M.D.Pi)& Path 2 &                        & -     &  \parbox{3cm}{termination\\(sole pathway)} \\ \hline
       \parbox{3.5cm}{(A.M.D, M.D.Pi)\\(A.M.D, M.D${}^*$)}
& Path 3 & term1                  & term2 & - \\ \hline
      (A.M, A.M.D.Pi)& -      & term2                  & -     & - \\ \hline
    \end{tabular}
  \end{minipage}
\end{table*}
  
Note that different models in the literature use quite different terminology
for dissociation pathways.  They are not entirely one-to-one translatable,
mainly because the models of Baker \cite{Baker.Warshaw2004} and Wu
\cite{Wu.Karplus2007} include weakly bound lead head states with ADP.Pi and ADP
on them, respectively.  Other models do not distinguish between weakly and
strongly bound states.  So Table \ref{tab:termination-dictionary} presents the
best effort of a ``dictionary'', showing equivalent dissociation (termination)
pathways in four models that include dissociation events.

With the parameters used in the original paper, the model reproduces realistic
run lengths for zero load.  It also shows a run length that falls with ATP for
concentrations above $50\,{\rm \mu M}$.  The run length decreases with an
increasing ADP concentration - however, the findings of Baker \emph{et al.}\
\cite{Baker.Warshaw2004} are only partially reproduced.  A more serious
discrepancy is that the dissociation rate becomes very high for strong forward
loads.  Newer experiments have shown that myosin V stays attached to actin for
a few seconds even under 10pN forward load \cite{Gebhardt.Rief2006}.
Therefore, dissociation pathways 1 and 2 are likely overestimated in
Ref.~\cite{Vilfan2005}.

The model was later expanded to take into account torsional fluctuations in the
actin structure, which lead to an improved agreement between the calculated and
the measured step size distribution \cite{vilfan2005b}. It was also used to
calculate the behaviour of myosin V when it encounters an Arp2/3 mediated
filament branch \cite{vilfan2008a}.  The predicted branching probability is in
good agreement with data by Ali \emph{et al.}\ \cite{Ali.Warshaw2007}.

\section{Conclusions}

Our comparison has shown that there is no myosin V model so far that would fit
\emph{all} available experimental data.  A particular problem is to explain the
high processivity over a wide range of loads.  Additionally, our experimental
knowledge about processivity is still somewhat incomplete.  Data on load
dependence of the run length for different ATP and ADP concentrations would be
immensely useful.  However, a glance at Table \ref{tab:termination-dictionary}
possibly reveals the lines of a future consensus on the dissociation mechanism.
Baker \emph{et al.}\ \cite{Baker.Warshaw2004} conclude that ``term1'' is the
predominant dissociation pathway.  In the elastic lever arm model
\cite{Vilfan2005}, three pathways are present, but only Path 3 shows a
dependence that allows a high degree of processivity for forward loads.  Wu
\emph{et al.}\ \cite{Wu.Karplus2007} consider two pathways, but use a higher
dissociation rate for ``term2''.  These terms all denote more or less the same
dissociation pathway, namely dissociation from the waiting, (A.M.ADP,A.M.ADP)
state.  The model by Skau \emph{et al.}\ \cite{Skau.Turner2006}, however, uses
a completely different path.  The consequence is that the run length strongly
decreases with load, so that the model motor loses processivity when the load
approaches 1pN.

With regard to sub-steps, there is still some discrepancy among experimental
results \cite{Uemura.Ishiwata2004,Cappello.Prost2007}.  This controversy will
have to be sorted out.  In addition, more data on load- and nucleotide
dependence of the substep size and duration will provide valuable input for
future theoretical models.

Finally, let us remark that all theoretical models described here are largely
phenomenological.  Connecting them in a quantitative manner with our knowledge
of the conformational changes that take place inside the myosin head, such as
opening/closing of switch 1, switch 2 and the cleft
\cite{Houdusse.Sweeney2001,Coureux.Houdusse2003,Coureux.Houdusse2004,Yengo.Sweeney2004,Holmes.Schroder2003,Conibear.Malnasi-Csizmadia2003,Reubold.Manstein2003},
remains a major challenge for the future.

\section{Acknowledgments}
I would like to thank Martin Karplus and Anatoly Kolomeisky for helpful
comments on the manuscript.  This work was supported by the Slovenian Research
Agency (Grant P1-0099).
\addtolength{\parskip}{-0.3cm}

\end{document}